# Error-Control and Digitalization Concepts for Chemical and Biomolecular Information Processing Systems

Vladimir Privman[*]

*Department of Physics, Clarkson University, Potsdam NY 13699, USA*

___________________________________

[*] Phone: +1-315-268-3891; E-mail: privman@clarkson.edu

## ABSTRACT

We consider approaches for controlling the buildup of noise by design of gates for chemical and biomolecular computing, in order to realize stable, scalable networks for multi-step information processing. Solvable rate-equation models are introduced and used to illustrate several recently developed concepts and methodologies. We also outline future challenges and possible research directions.

# 1.  Introduction

There have been significant advances in the development of chemical[1-4] and biomolecular[5-12] systems which are intended to process information by realizing Boolean gate functions, for example, **AND**, **OR**, etc., in (bio)chemical kinetics. The range of the information carrying entities has not been limited to simple molecules, but included[1-13] supra-molecular and biomolecular structures (enzymes, DNA, etc.), as well as whole cells. Most expected uses of chemical or biochemical "information processing", to be termed "computing" for brevity, have not aimed at replacing the conventional computers but rather at offering additional functionalities for multi-signal sensing[14,15] and interfacing/actuation[15-17] in situations when a direct wiring to computers and power sources is not practical, such as in many biomedical applications.

One of the main challenges for chemical and biochemical computing has been design of gates and other processing elements, with capabilities to connect them as network components for fault-tolerant information processing of increasing complexity.[18-20] First results in developing networks for (bio)chemical information processing[1-4,18-20] have included diverse systems, such as those performing elements of basic arithmetic operations,[21-22] multifunctional molecules,[23-25] DNA-based gates and circuits,[26,27] and enzyme-catalyzed reaction networks of several concatenated gates.[19,20,28,29]

In this article we survey selected topics in, as well as offer illustrative model examples of theoretical analyses of noise reduction and control for scalability in biochemical computing, recently developed by our group[10,12,14,17,19,30] primarily in conjunction with experimental data for enzyme-reaction based logic gates and networks. Theoretical studies, which generally apply to a broad range of chemical and biomolecular information processing systems, presently suggest that typical networks up to 10 gates can operate with the acceptable level of noise,[10,12,17] similar to recent findings in networking of neurons.[31,32] For larger networks, additional non-Boolean network elements, as well as proper network design to utilized redundancy for digital error correction will be needed for fault-tolerant operation.[10,12,17,30]

There is plentiful experimental evidence that the level of noise in chemical[1-4] and biomolecular[5-12,14,17,19-20,31-34] computing systems is quite high as compared to the electronic computer counterparts. This includes both the input/output signals and the "gate machinery" chemical concentrations, which typically vary at least several percent on the scale normalized to the digital **0** to **1** range. Avoiding noise amplification by gate and network design is therefore quite important even for rather small networks. While we consider aspects of error control, here we do not address the *origin/sources* of stochastic noise in (bio)chemical reactions: This would take us into topics in statistical mechanics which are outside the scope of this article. We also do not review the experimental findings, which are illustrated in other articles in this Special Issue. Instead, here we devise solvable chemical rate equation models and use the resulting expressions to illustrate recently developed concepts in (bio)chemical computing gate design for noise control and suppression.

Let us further discuss the plethora of chemical and biomolecular information processing systems. In the remainder of this section and in the next section, most literature citations aim at providing examples and are not exhaustive. We first reference an extensive body of ongoing experimental work on chemical processes reformulated[35,36] in the language of computing operations, involving changes[1-4,36-48] in various structural, chemical, or physical properties upon application of physical,[49-76] chemical,[77-84] or more than one type[85-87] of input signals. The final output signals have typically been read out spectroscopically[88-93] or electrically/electrochemically.[94-96] Chemical computing can be done in a bulk system, specifically, in solution,[97-98] or at surfaces/interfaces,[14-17,99-102] such as electrodes or Si-chips. Supra-molecular ensembles can also operate as switchable "molecular machines" performing logic operations.[103]

Much effort has been invested in realizing chemical-computing equivalents of standard Boolean gate functions, such as **AND**,[104,105] **OR**,[106] **XOR**,[103,107] **NOR**,[108-111] **NAND**,[112,113] **INHIB**,[114-117] **XNOR**.[118-119] Issues of reversibility,[120,121] reconfiguring[122-125] and resetting[126,127] logic gates have been explored. Somewhat more complicated systems[128-134] have carried out digital logic functions of several-gate rudimentary device components (e.g., keypad lock) and memory units.[135-145] Chemical-computing systems can encode computational steps at the single-molecule[146] "nano-scale" level,[147] as well as perform parallel computations by numerous molecules.[148]

In summary, chemical computing shows great promise.[149-151] However, as most unconventional computing paradigms,[152] it is not considered a foreseeable-future viable alternative to the speed and versatility of Si-chip computers. Rather, it offers new functionalities and novel approaches to applications, such as microrobotics, multi-input (bio)sensors/actuators, and bioimplantable devices. As with other unconventional computing systems, the main challenge for chemical computing has been networking of basic gates for achieving scalable, fault-tolerant information processing similar to "ordinary" electronic computers.[153] Advances have recently been reported with small networks performing basic operations,[1-4,21,22] for example, adder/subtractor and their sub-units.[154-159] Multi-signal response to chemical or physical inputs has been explored,[23-25] and attempts at scaling up the complexity of chemical computing along the lines of biological principles have been made.[74]

Most variants of biomolecular or *biochemical* information processing, to be termed "biocomputing" for brevity, can be considered a branch of chemical computing. However, it has developed into an independent, active field of research. The reasons have been several-fold. Indeed, biomolecules offer natural specificity when used in complex "chemical soup" environments, as well as biocompatibility, the latter important for biomedical and biotechnology applications. Biomolecules are also likely more suitable for possible future developments of scalability paradigms borrowing ideas from Nature. Furthermore, the biocatalytic nature of many biomolecular processes, offers certain advantages for analog noise control within the Boolean-gate circuit design paradigm.[17] Biomolecules such as proteins/enzymes[5-12,14-17,19-20,161-165] and DNA/RNA/DNAzymes[166-182] have been extensively used for biocomputing, including systems realizing small networks and computational units, and those motivated[183] by applications.

This article is organized as follows. In Section 2, we introduce general concepts for considering (bio)chemical gate functions. Gate design for decreasing the degree of noise amplification is addressed in Section 3. Section 4 presents optimization of AND gates, while Section 5 describes gate design as part of a small network. Section 6 offers a summary and also discusses future challenges.

## 2. Analog/Digital Paradigm for (Bio)chemical Information Processing

In order to realize networks processing large quantities of information at high levels of complexity, the approach envisioned in the chemical and biochemical computing literature has usually been that gates will be connected similar to Si-chip electronic devices, paralleling the "conventional" design[184,185] of fault-tolerant systems that can avoid buildup of noise without the prohibitive use of resources. Another approach, particularly with biomolecules involved, could be the use of design concepts borrowed from processes in living organisms, which are being actively studied in Systems Biology.[186] Ultimately, hybrid solutions can be expected, with bio-inspired elements supplementing the conventional design. One can also utilize massive parallelism,[182] such as in variants of DNA computing.[187]

There are several good reasons for the enzyme-based biocomputing gate and network realizations and analyses reported by our group,[10,12,14-17,19,20,30] and for much of the rest of the biomolecular computing literature, to follow the conventional information processing paradigm of modern electronics: digital approach based on analog gates and other elements operating in a network.[184,185] Indeed, biomolecular computing is presently far from the complexity and richness of coupled biochemical reaction sets needed for mimicking processes in living organisms. Furthermore, most applications of the near-future moderate-complexity biocomputing systems are expected to be for novel sensor development,[15] involving processing of several input signals and yielding Yes/No digital outputs, corresponding to "Sense/Respond" or "Sense/Diagnose/Treat" actions. Thus, either in the biochemical stages or during signal transduction to electrodes/electronic computers for the "action" step, the Yes/No digitalization will be imposed, for example, by filtering, as addressed later.

Most importantly, the use of the digital information processing paradigm offers a well established approach for control of the level of noise buildup in networks of (bio)chemical information processing reactions. Chemical and biochemical systems are much more prone to noise than electronic computers. The inputs reactant concentrations, and the "gate machinery" chemical concentrations, such as those of catalysts, are expected to fluctuate within at least a couple of percent of the range of values between the "digital" **0** and **1**. As a result, consideration of control of noise build-up is needed already for small, 2-3 gate networks.[10,15,19,20]

Digital information processing is actually carried out by network elements which are analog in nature. Figure 1 illustrates the simplest "gate": the identify function. A possible analog response curve is also shown. The input and output signals are not limited to the range bounded by the selected "digital" **0** and **1** values appropriate for a specific application. They can also be considered for values beyond the "digital" range, if physically allowed (for example, chemical concentrations can only be nonnegative), as illustrated by the broken-line sections in the figure. The "digital" **0** does not have to be at the physical zero.

Let us consider a simple model for a chemical reaction, described within the rate-equation approximation, of two atoms of the species $A$, of initial concentration $A(0) = A_0$, combining to yield the product, $C$, of concentration $C(t)$ at time $t \geq 0$, where initially, $C(0) = 0$:

$$A + A \xrightarrow{k} C. \tag{1}$$

Here $k$ denotes the rate constant for the reaction, assumed irreversible. The rate equation is easily solved:

$$\frac{dA}{dt} = -2kA^2, \tag{2}$$

$$C(t) = \frac{A_0 - A(t)}{2} = \frac{kA_0^2 t}{1 + 2kA_0 t}. \tag{3}$$

We will further assume that the information-processing application involves a certain value, $A_{max}$ of the input which is regarded as the digital **1**, and, for simplicity, we also take the physical zero as the digital **0** input. We will also assume that the product of the reaction provides the output signal at the "gate time" time $t_g$. It transpires that the digital value for the output is then set by the gate/application itself and cannot be conveniently selected. Digital **0** and **1** will be at, respectively, $C = 0$ and

$$C_{max} = \frac{kA_{max}^2 t_g}{1 + 2kA_{max} t_g}. \tag{4}$$

Next, we consider logic-range variables, in terms of which the input, $A(0) = A_0$, and the output, $C(t_g)$, are normalized to the "digital" range of values:

$$x = A_0 / A_{max}, \quad (5)$$

$$z = C(t_g)/C_{max}. \quad (6)$$

From these relations, we can get the gate-response function shape, see Figure 2,

$$z(x) = \frac{(1+2p)x^2}{1+2px}, \quad (7)$$

which depends on the combination of parameters:

$$p = kA_{max}t_g. \quad (8)$$

It is interesting to note that, while the digital-**1** of the input, $A_{max}$, is generally set by the application, we can control the reaction rate constant, $k$, by varying the physical and chemical conditions of the system to the extent allowed by the application. We can also adjust the reaction duration, $t_g$. Thus, there is a certain degree of control of the "response function shape" which could be used for gate design and optimization. To discuss this further, we have to address the issue of control of noise, which is the topic of the following sections.

Here we point out that the considered chemical reaction generally can only yield concave shapes of the type shown in Figure 2. However, as seen in the figure, even with this limitation the shape of the gate response does not vary significantly. Large variations in the parameter values are needed to achieve qualitatively different response. This difficulty has been noted and discussed earlier[10,12,19] and is shared by most biocatalytic information processing systems which realize specific gates studied for novel sensor applications.[15] Finally, we note that the ranges of both variables in Figure 2 need not be limited to [0,1]; they can be considered for $x$ and $z$ larger than 1 as well.

## 3. Noise Amplification and Filtering

In order to discuss noise amplification and filtering, we will, in this section, continue to consider the simplest "identity" gate as a reference. Two-input/one-output gates, such as **AND**, will be addressed later. However, let us introduce a different reaction which offers a response more realistic of typical chemical kinetics. We consider the process

$$A + B \xrightarrow{K} C, \qquad (9)$$

with the rate constant $K$ and initial conditions $A(0) = A_0$, $B(0) = B_0$, $C(0) = 0$, and the output signal again measured as $C(t_g)$. Obviously, this can also be perceived as a two-input **AND** gate. However, here we prefer to regard $A_0$ as the input set by the application (the environment in which the gate is used), whereas $B_0$ $(< A_0)$ will be assumed small (so that it will limit rather than drive the output) and regarded as a controllable-supply "gate machinery" chemical.

The rate equation is then

$$\frac{dA}{dt} = -KAB = -KA(A - A_0 + B_0), \qquad (10)$$

yielding

$$C(t) = \frac{A_0 B_0 [1 - e^{-(A_0 - B_0)Kt}]}{A_0 - B_0 e^{-(A_0 - B_0)Kt}}. \qquad (11)$$

Equations (5)-(6) are then used to rescale the input and output in terms of the "logic" ranges, with the result

$$z(x) = \frac{x(1 - e^{-ax+b})(a - be^{-a+b})}{(1 - e^{-a+b})(ax - be^{-ax+b})}, \qquad (12)$$

which depends on the two combinations of parameters:

$$a = KA_{\max}t_g, \quad b = KB_0 t_g. \qquad (13)$$

As before, these parameters can be controlled by changing the physical and chemical conditions (vary $K$), the "gate machinery" chemical supply, $B_0$, and the reaction time, $t_g$, as illustrated in Figure 3.

An important observation, which can be proven by tedious algebraic considerations not reproduced here, is that the function in Equation (12) always gives the monotonically increasing, convex response curve; see Figure 3. Typically, in catalytic biochemical reactions such convex response curves (and surfaces, for more than one input) are also found: The output (the product of the reaction) is controlled by and is typically proportional to the input-signal chemical concentration(s) for small inputs. For large inputs, the output is usually limited, for example, by the reactivity of the available biocatalyst, and the response signal reaches saturation.

We point out that the "digital" **0** and **1** signal values need not be sharply defined. In some applications, input or output signals in certain ranges of values may constitute **0** or **1**. For example, a certain range of "normal" physiological concentrations can be **0**, whereas another, "abnormal" range, can be **1**. These ranges need not even be bounded, for instance, if the "abnormal" concentrations correspond to those above a certain "flag" value.

There are several sources of error in gate functioning. The most obvious source of noise is that in the inputs, which is natural and actually quite large in chemical and biochemical environments in which applications of (bio)chemical information processing have been envisioned. The gate function will transfer this noise (the distribution of the input values) into noise in output signal(s).

The gate function itself can also be noisy and, furthermore, perhaps somewhat displaced away from the desired digital values/ranges. In our earlier examples, noise and fluctuations in concentrations and physical parameters of the system can lead to a distribution of the values of $z(x)$, for each $x$, rather than a sharply defined function such as in Equations (7), (12). Furthermore, the mean values of this distribution need not pass precisely through the expected logic values connected by smooth response surfaces. For example, the mean value of $z(1)$ might somewhat deviate from 1 for our "identity gate."

We will denote as "analog" the noise due to the spread of the output signal about the desired "digital" values (or possibly ranges of values). In order counteract buildup of noise, we have to pass our signals through "filters" of the type shown in Figure 4. In fact, ideally we would like to have the sigmoid property — small slopes/gradients at and around the digital points — in all or most of our gates. Filters can also be used as separate elements/steps. There is evidence that such solutions for suppressing analog noise buildup are utilized by Nature.[188,189]

However, filtering can push those values which are far away from the correct digital result (the values which are in the tail of the distribution about the desired correct outcome) to the wrong answer. Thus, the process of digitalization itself introduces also the "digital" type of noise. Such errors are not very probable and only become important to actively correct for larger networks. Standard techniques based on redundancy are available[190,191] for digital error correction.

For biocatalyst (typically, enzyme) based computing gates studied by our group, for the presently realized network sizes and levels of noise, it is the analog error correction that is important and has recently received significant attention.[10,12,14,15,17,19,20,30] It has been estimated[10,12] that up to order 10 such gates can be connected in a network before digital error correction is warranted.

Experimental realizations of the sigmoid behavior (Figure 4) are an ongoing effort and will hopefully be soon accomplished, along the lines of theoretical suggestions[10,189] based on the idea that an additional reactant which depletes the product, but can only consume (react with) a small quantity of it, will suppressed response at small inputs without voiding the saturation property at large inputs, thus yielding a sigmoid response. No simple solvable rate-equation models of the type analyzed earlier, can be offered here for sigmoid behavior: Our group's numerical simulations (work in progress) of rate equations for typical biocatalytic reactions support the proposed mechanism.

As pointed out earlier, the final output signal of several (bio)chemical information processing steps in near-term future sensor applications of the "decision-making" type,[14,15] will likely be coupled to conventional electronics. There are active, well developed research areas (not reviewed here) of interfacing enzyme-based logic with "smart" signal-responsive[16,192-205] materials and with electrodes and bioelectronic devices.[16,206-212] This interfacing, involving the transduction of (bio)chemical signals to electronic ones, can also involve a well-defined filtering "sigmoid" property, as has been recently experimentally demonstrated.[16]

## 4. The AND Gate

Logic **AND** gates are the most studied in chemical and biochemical computing, and the only ones explored in detail for the shape of their response surface, the latter in the literature on enzyme-catalyzed biochemical reactions, to be referenced later in this section. Indeed, the truth table for the **AND** gate is that the output **1** is obtained *only* when *both* inputs are **1**, which is the most natural outcome when measured as a product of a two-input chemical reaction. Presently, let us introduce a simple model for the **AND** gate chemical-computing function. For this, we now regard the reaction in Equation (9): $A + B \to C$, as a two-input, one-output process. We introduce the second variable reduced to the "logic" range [0,1],

$$y = B_0 / B_{\max}, \qquad (14)$$

to supplement the definitions in Equations (5)-(6), where $B_{\max}$ is the reference value of the logic-**1** for the second input, set by the application. We now regard the quantity $z$ defined in Equation (6), as a two-variable function, $z(x, y)$, describing the **AND**-gate response surface shape. The solution of the rate equations, given by Equation (11), is now recast in terms of the reduced variables to yield

$$z(x, y) = \frac{xy(\alpha e^\alpha - \beta e^\beta)(e^{\alpha x} - e^{\beta y})}{(e^\alpha - e^\beta)(\alpha x e^{\alpha x} - \beta y e^{\beta y})}, \qquad (15)$$

where we defined the parameters

$$\alpha = KA_{\max}t_g, \quad \beta = KB_{\max}t_g. \tag{16}$$

This is similar to the set of (dimensionless) parameters in Equation (13). However, here we have less control over their values, because their ratio is fixed by the application (the environment) of the gate which in many cases dictates the values of $A_{\max}$ and $B_{\max}$. Thus, technically only the product $Kt_g$ can be adjusted.

The shapes of the resulting response surfaces are illustrated in Figures 5 and 6. As mentioned earlier, the noise in the input signals is passed on to the output, with the added noise effects due to the gate functioning itself, such as the imprecise (on average) and fluctuating values of $z(x,y)$. In addition to designing gates with as precise and sharply defined $z(x,y)$ as possible, we can also minimize the propagation of analog noise, and hopefully avoid noise amplification, by finding parameter (such as $Kt_g$) values that yield gates which suppress spread in the input signals by having small slopes near the logic points. Let us consider this concept in greater detail.

The absolute value of the gradient vector, $|\vec{\nabla}z(x,y)|$, calculated at the logic points, measures the noise spread amplification or suppression. However, this is only relevant provided the gate function $z(x,y)$ is smooth (does not vary much) in regions about the logic points which are approximately the size of the spread of the noise distributions. Our model example offers illustration of relatively smooth $z(x,y)$ shapes; see Figure 5. We can try to identify parameter values for which the *largest* of the four gradients, $|\vec{\nabla}z(x,y)|_{x=0,y=0}$, $|\vec{\nabla}z(x,y)|_{x=1,y=0}$, $|\vec{\nabla}z(x,y)|_{x=0,y=1}$, $|\vec{\nabla}z(x,y)|_{x=1,y=1}$ is as small as possible (note that $|\vec{\nabla}z|_{00}$ is always zero for this particular model). For this calculation, we will assume that both $\alpha$ and $\beta$ can be adjusted independently, which gives additional freedom (we commented earlier that in applications the ratio of these two parameters might be fixed and not controllable). What we are after is an estimate of how little can noise amplification be for **AND** gates modeled by this reaction scheme, $A+B \rightarrow C$. By numerical calculation, we find that for moderate values of $\alpha$ and $\beta$, the minimum is obtained for $\alpha = \beta \approx 0.4966$, and is given by $|\vec{\nabla}z|_{10} = |\vec{\nabla}z|_{01} = |\vec{\nabla}z|_{11} \approx 1.1796$.

This result is interesting in several aspects. First, gate functions of this type amplify analog noise even under optimal conditions. The noise amplification in the best case scenario is about 18%. Studies[10,12,19] of enzyme-based **AND** gates, which have utilized more realistic (and thus more complicated and not exactly solvable) rate-equation models appropriate for biocatalytic reactions, found similar estimates. Experimental data were fitted and results were numerically analyzed by using both the rate equation approach and more phenomenological shape-fitting forms for the gate response function surface, the latter described in the next section. It transpires that smooth, convex gates corresponding to (bio)chemical reactions can have very large noise amplification, typically 300–500%, if the gate is not optimized. Reaching the optimal conditions is not always straightforward primarily because the gate function shape depends only weakly on parameter values. Even under optimal conditions, at least about 20% noise amplification is to be expected.

For fast reactions, the maximum of the four gradients can actually be smaller than ~1.18, and in fact can even be somewhat less than 1 (which would suggest noise suppression). However, as seen in Figure 6, under these conditions the gate function surface develops sharp features, and the gradients can no longer be used, because they remain close to the logic-point values only in tiny regions near these points, as compared to the typical noise spread of at least several percent, for (bio)chemical signals.

For such gates, and generally when the spread of the noise is larger than the $x$ and $y$ scales over which the gate function varies significantly, one can assume a certain shape of the noise distribution, such as a product of approximately Gaussian (half-Gaussian, if the logic zero is exactly at the physical zero) distributions in $x$ and $y$, for inputs at each of the logic points. When this distribution is properly integrated with the use of the gate response function,[10] one can numerically calculate the output signal distribution for each of the input options, and thus estimate noise propagation.[10,12,14]

Interestingly, a "ridged" gate response function was encountered[12] in a study of an enzymatic system, which has also a smooth-response counterpart when a different chemical is used as one of the inputs.[12] While the reaction kinetics was much more complicated than the present model, the finding has been quite general for such gate functions: The optimal conditions are obtained with a symmetrically (diagonally) positioned ridge, as in Figure 6, and noise amplification percentage is then very small (estimated not from gradients, but by integrating over distributions). For gates operating in this regime, with the amplification factor only slightly larger than 1, noise amplification is practically avoided (when compared to other possible noises of source, from the gate-function itself). However, they do not have the noise-suppression, "filter" property.

Figure 7 offers a schematic of another **AND** gate-response shape which was recently explored and experimentally realized: sigmoid in only one of the two inputs. Many allosteric enzymes have such a "self-promoter" property with respect to one of their substrates (input chemical species that the enzyme binds as part of its biocatalytic reaction scheme). Details[14] for this specific experiment and its modeling are not given here. A key finding[14] was that the single-sided sigmoid shape can be tuned by parameter adjustment to also have the noise amplification percentage very small (noise amplification factor only slightly above 1), so that there is practically no noise amplification. However, a desirable two-sided sigmoid response, also illustrated in Figure 7, has not been to our knowledge realized at the level of a single **AND** gate, in chemical or biomolecular computing literature. Certain biochemical processes in Nature, which are much more complex than our synthetic **AND**-gate systems, do realize[188] a two-sided sigmoid response.

## 5.  Network of AND Gates

Optimization of (bio)chemical gates one at a time is not straightforward for several reasons. Indeed, we have seen that in most cases a rather large variation of the controllable parameters is needed: physical and chemical conditions, reactant concentrations and in some cases choice of species, which may not be experimentally feasible. In fact, the actual detailed kinetic modeling of the reactions involved, especially for biomolecular systems, is in itself a challenging and numerically taxing task, not reviewed here.[10,12,14,15,17] Furthermore, the kinetics of most biomolecular processes, specifically those used for **AND** gates, is not only complex but also not well studied. The quality of the experimental data for the gate-response function shape is limited due to the noise in the gate-function itself, short life-time for constant activity of the biocatalytic species, etc. As a result, multi-parameter complex reaction schemes, even if proposed, are difficult to substantiate by data fitting in the gate-design context which requires models to work for a large range of parameters.

It is therefore useful to explore optimization of the relative gate functioning as part of a network, whereby each gate is modeled within an approximate, phenomenological curve/surface-fitting approach. Such ideas have recently been tested[19] for coupled enzymatic reactions which include steps common in sensor development[213] for maltose and its sources. A modular network representation of the biocatalytic processes involved is possible in terms of three **AND** gates; see Figure 8. This convenient representation is actually approximate, because it obscures some of the complexity of the processes involved, which are not reviewed here.[19]

The approach taken, has been as follows. We first propose an approximate, phenomenological fitting function for the gate response surface in terms of as few parameters as possible, but enough to capture the expected global, qualitative features of the shape. Specifically, for a typical convex "identity" gate, the proposed fitting function is here conveniently written as

$$z(x) \approx \frac{sx}{(s-1)x+1}. \qquad (17)$$

This is just a simple, single-parameter, $s$, rational form that "looks" qualitatively appropriate, provided we assume that

$$s > 1. \qquad (18)$$

Indeed, the curve is then convex and has slope $s$ at $x, z(x) = 0, 0$, and $1/s$ at $x, z(x) = 1, 1$.

For a convex, smooth **AND** gate, we use the two-parameter, say, $s > 1$ and $u > 1$, product function,

$$z(x, y) \approx \frac{(sx)(uy)}{[(s-1)x+1][(u-1)y+1]}. \qquad (19)$$

The gradient values are $|\vec{\nabla}z|_{00} = 0$, $|\vec{\nabla}z|_{10} = u$, $|\vec{\nabla}z|_{01} = s$, and $|\vec{\nabla}z|_{11} = \sqrt{s^{-2}+u^{-2}}$. The minimum of the largest of the last three values is obtained for $s = u = \sqrt[4]{2} \approx 1.189$, which is also the *value* of the gradient, consistent with the earlier reported expectation that smooth convex **AND** gates can typically be optimized at best to yield noise amplification somewhat under 20%.

Having introduced our approximate fitting functions, we now vary selective inputs in the network; see Figure 8. In the experiment,[19] each of the three inputs $x_{1,2,3}$ was separately varied between 0 (corresponding to the logic **0**) and the reference value pre-defined as the logic **1**, while all the other inputs (including $y_3$) where at their reference logic-**1** values. In fact, when the parameterization of Equation (19) is applied to all three gates in our network, Figure 8, we get a complicated rational expression for $z$ as a function of all the four inputs ($x_{1,2,3}$ and $y_3$). Setting all of them but a single *x*-input to 1, we get the parameterization for the measurement with that input varied. We only keep the varying arguments for simplicity:

$$z(x_1) = \frac{s_1 x_1}{(s_1 - 1)x_1 + 1}, \qquad (20)$$

$$z(x_2) = \frac{s_2 u_1 x_2}{(s_2 u_1 - 1)x_2 + 1}, \qquad (21)$$

$$z(x_3) = \frac{s_3 u_1 u_2 x_3}{(s_3 u_1 u_2 - 1)x_3 + 1}. \qquad (22)$$

An interesting conclusion is that each data set only depends on a single parameter ($s_1$, or one of the products $s_2 u_1$ or $s_3 u_1 u_2$).

Thus, we only get partial information on the gate functioning. However, we can attempt to "tweak" the relative gate activities in the network to improve its stability. We note that if the proposed approximate description is accurate for a given gate, then the parameters $s$ and $u$ for that gate will be functions of adjustable quantities, such as the gate time, input concentrations of some of the chemicals, reaction rates (which can in turn be controlled by the physical and chemical conditions). In addition, $s$ and $u$ can depend on other quantities which are not controllable in a specific application. Without detailed rate-equation kinetic modeling and verification of applicability of the phenomenological functional form selected, this parameter dependence is not known.

However, examination of the fitted quantities ($s_1$, $s_2 u_1$, $s_3 u_1 u_2$) still provides useful information on the relative effect that the gates have in their contribution to the gradients at various logic points, when compared to the optimal values ($s_1 = 2^{1/4}$, $s_2 u_1 = 2^{1/2}$, $s_3 u_1 u_2 = 2^{3/4}$). The initial sets of data[19] were collected with the experimentally convenient but otherwise initially randomly selected values for the adjustable "gate machinery" and other parameters. Examination of the results[19] has lead to a semi-quantitative conclusion that the deviations form the optimal values could largely by attributed to the gate which is the closest to the output in Figure 8 ($z = x_1$ **AND** $y_1$): it was too "active" as compared to the other two gates (means, its biocatalytic reaction was too fast). A new experiment was then carried out[19] with the concentration of the enzyme catalyzing this gate's reactions reduced by an order of magnitude (actually, by a factor of approximately 11); recall that large parameter changes are needed to effect qualitative changes. The new data collected for the modified network yielded $s_1$, $s_2 u_1$, $s_3 u_1 u_2$ values significantly closer to optimal.[19]

## 6.  Conclusions and Future Challenges

We addressed certain aspects of and approaches to gate optimization for control of the analog noise buildup, which is an important consideration in connecting gates in functional networks, though for larger networks digital error correction by redundancy will also have to be implemented, and special network elements will have to be devised, notably filters, but also elements for signal splitting, balancing and gate-to-gate connectivity, memory, and interfacing with external input, output and control mechanisms.

Our goal here has been to develop simple rate-equation models which allow exact solvability, and then use them to illustrate and motivate the discussion. Thus, we avoided presenting experimental data and their analysis, which can be found in the cited articles, while various chemical and biochemical gate examples are offered in other reviews in this Special Issue.

The reader will notice that our presentation has been limited to **AND** gates and related systems. The reason for this has been that all the recent studies of noise control in (bio)chemical computing have thus far been for **AND** gates and, furthermore, only those with the logic **0** set at the physical zeros of chemical concentrations. While such a limitation is natural for chemical reactions per se, it is definitely not typical for applications envisaged, especially in multi-input biomedical sensing.[15]

We expect that, as new experiments on mapping out (bio)chemical gate functions and probing network functioning are reported (some presently ongoing in our group), new features in noise and error control will be explored. Specifically, noise in the gate function itself, including spread of its values and imprecise mean-value — not exactly at the reference **0** or **1**, with deviations possibly also somewhat different for various inputs that should ideally yield the same logic output — will have to be considered and corrected, most likely by filtering. Indeed, we expect that while long-term network design and scaling up will be crucial, the grand challenge in (bio)chemical information processing short-term is to design versatile and effective (bio)chemical filter processes that can be concatenated with various types of single logic gates.

The author gratefully acknowledges research funding by the NSF (grant CCF-0726698) and by the Semiconductor Research Corporation (award 2008-RJ-1839G).

# FIGURES

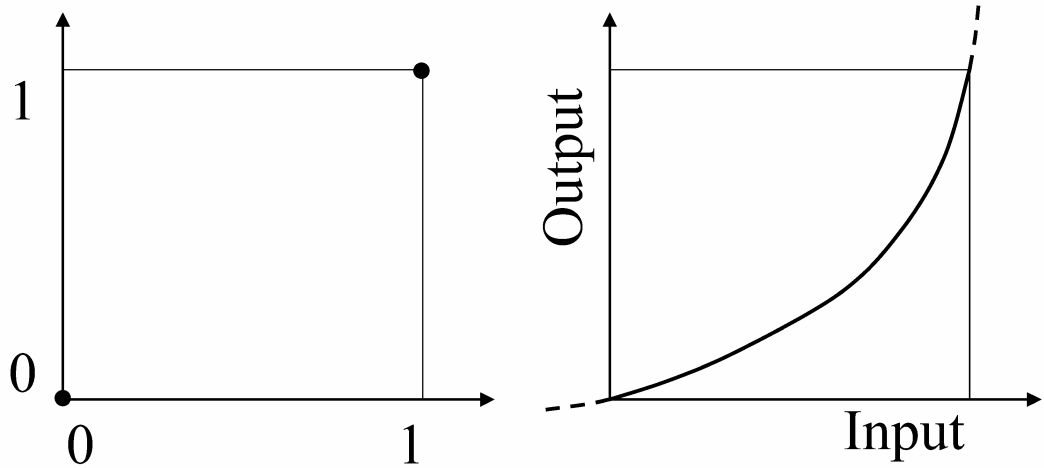

**Figure 1.** *Left:* The identity "gate" mapping digital **0** and **1** to the same values. *Right:* A possible response curve.

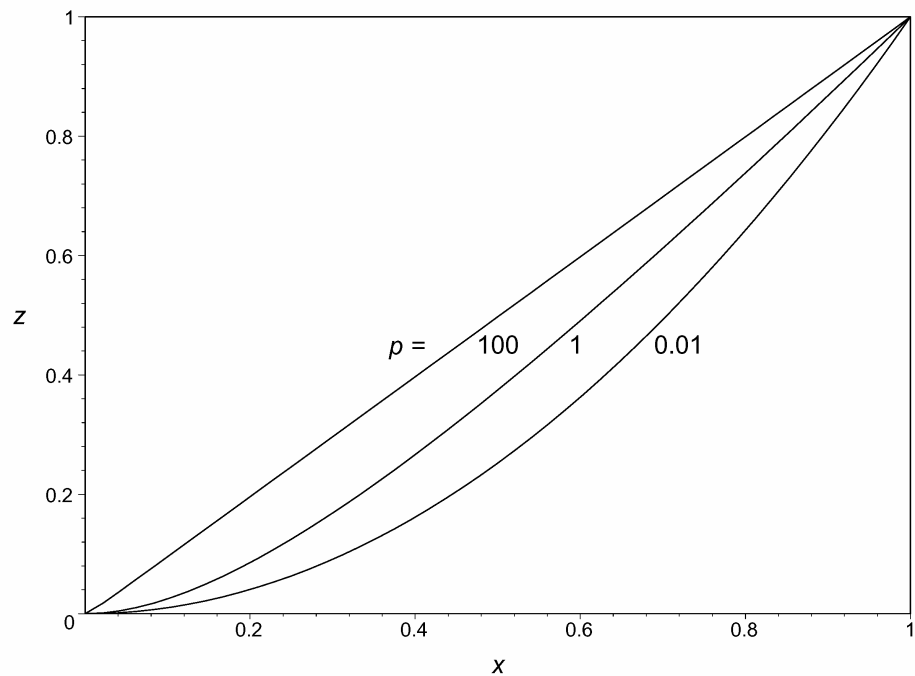

**Figure 2.** The response function corresponding to the reaction $A + A \to C$, see Equation (7), for three different values of the parameter $p$. All three curves are concave.

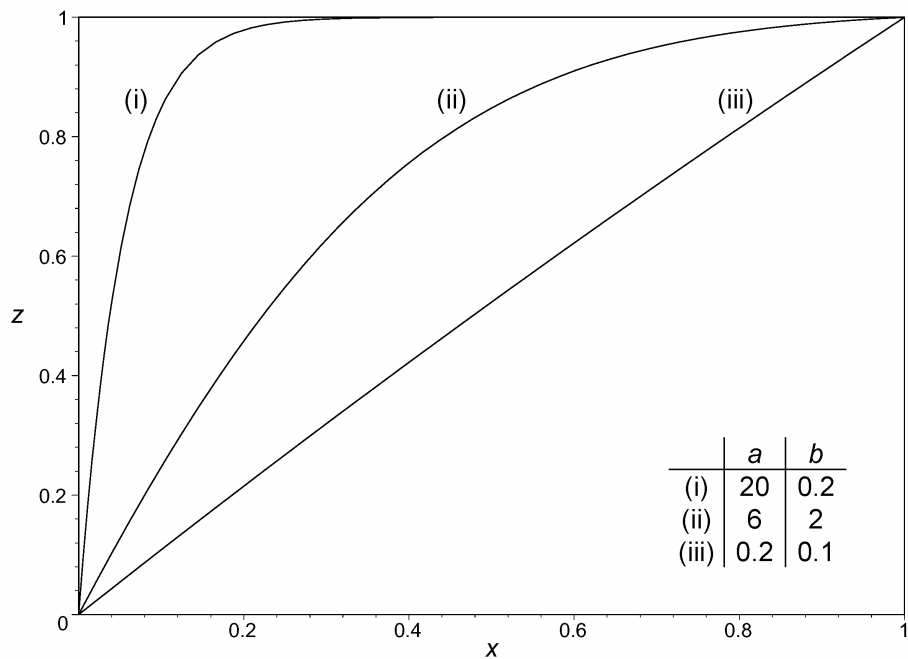

**Figure 3.** The response function corresponding to the reaction $A + \cdots \to C$, where the omitted reactant is not considered a variable input, see Equation (12), for three different pairs of values of the parameters *a* and *b*. All three curves are convex.

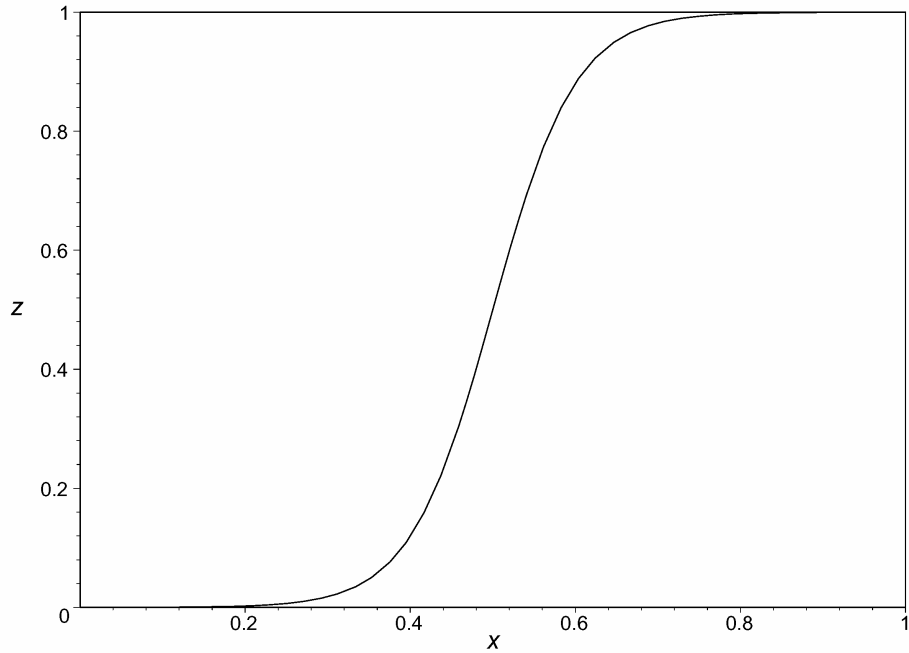

**Figure 4.** A desirable sigmoid response for filtering: the central inflection region is narrow and positioned away from both logic **0** and **1**, and the slopes at and near both logic values are very small (ideally, they should be zero).

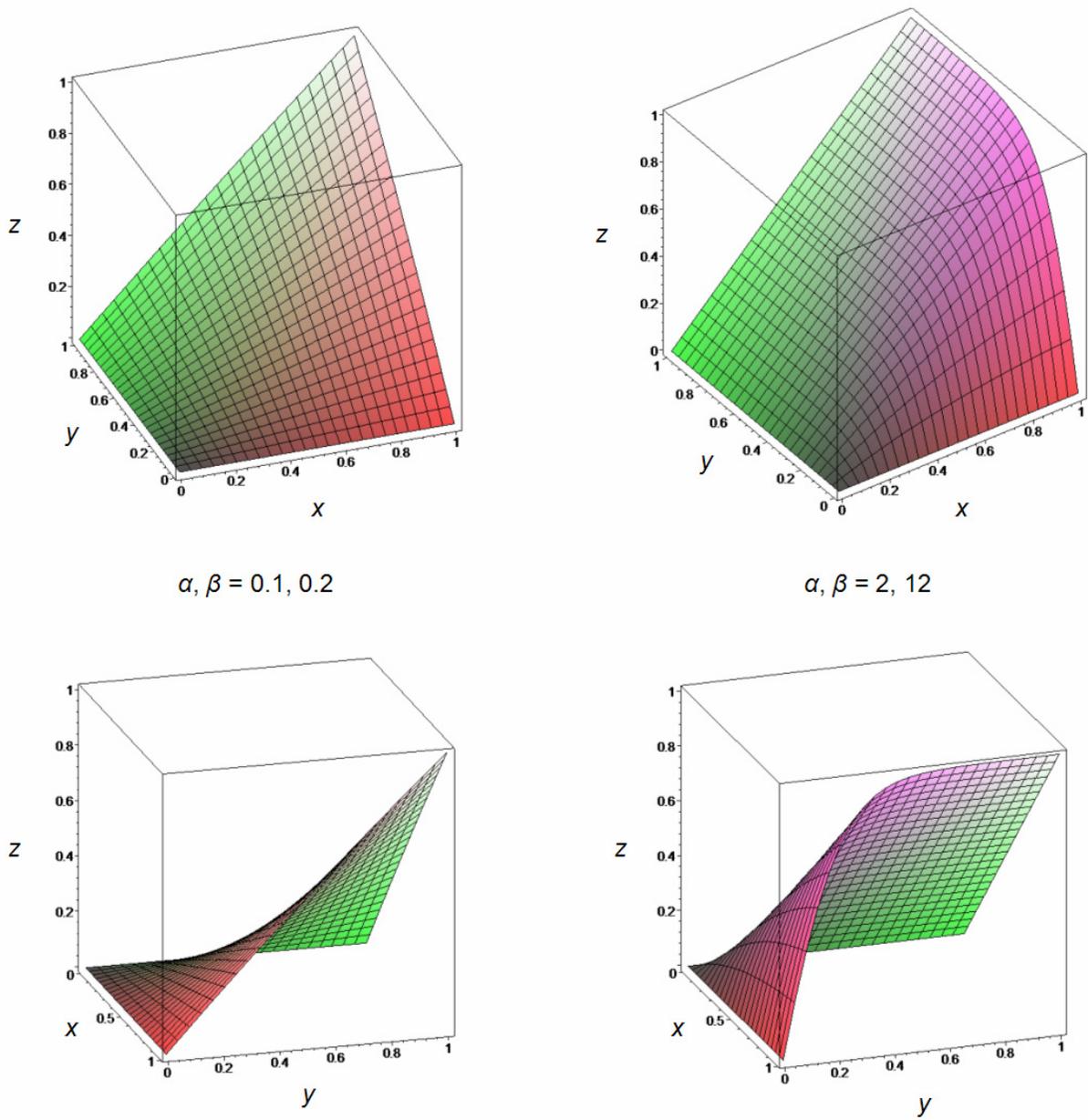

**Figure 5.** Smoothly varying response surfaces of the **AND** gate realized by the reaction $A + B \to C$, see Equation (15), for two choices of the parameters $\alpha$ and $\beta$ defined in Equation (16). The upper panels give the front view, whereas the lower panels offer the back view of the same surfaces.

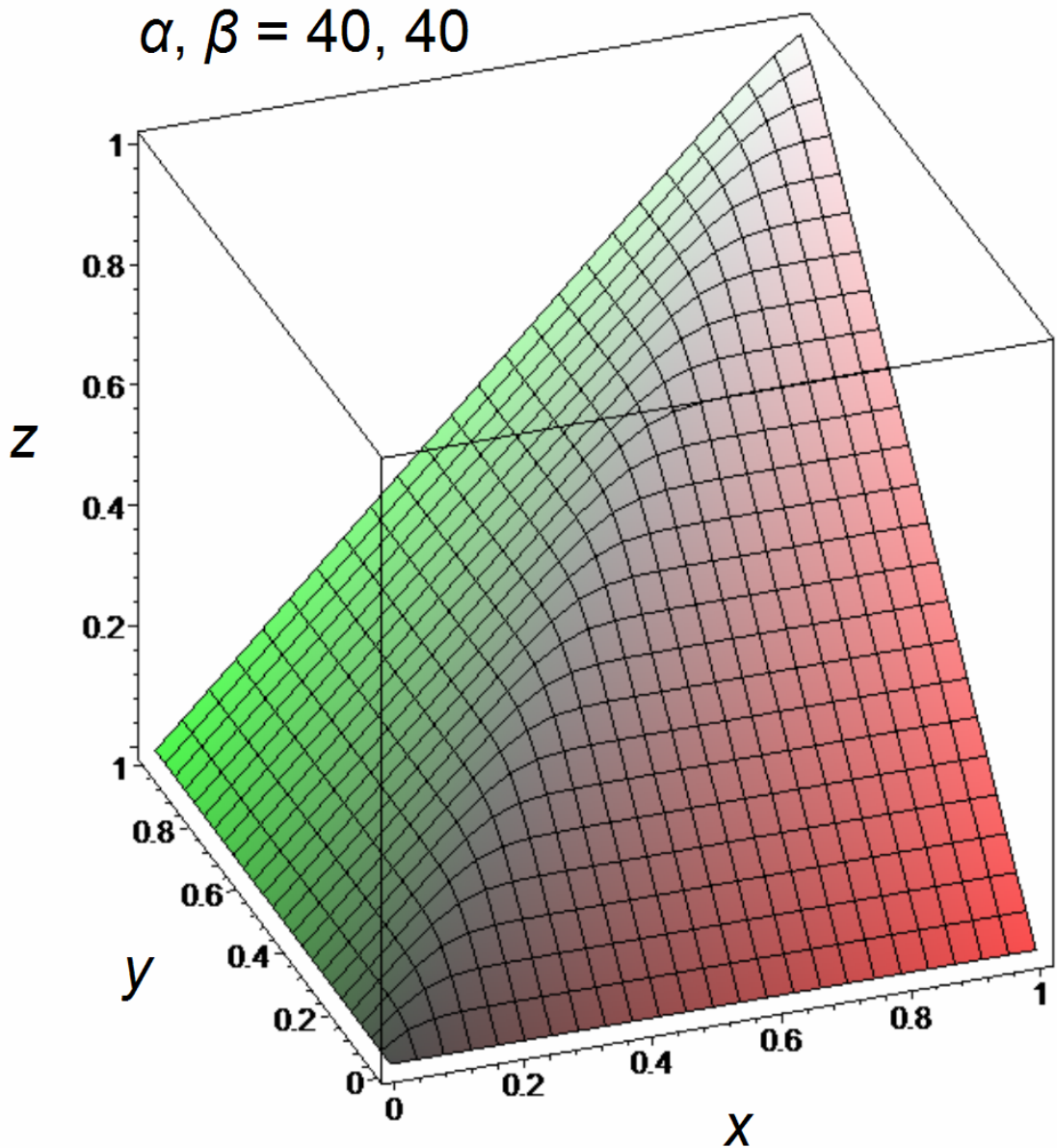

**Figure 6.** Same as in Figure 5, but with fast-reaction parameters (large $\alpha$ and $\beta$). This case illustrates the emergence of a response surface with non-smooth features: formation of a ridge (here symmetric, along the diagonal), and also shrinking of the region for which the value of the gradient near the point (0,0) remains small. Note that the gradient *at* the origin, $\left|\vec{\nabla}z(x,y)\right|_{x=0, y=0}$, is zero for all the surfaces shown in both figures. Similar nonuniformities set in all along the ridge region, including near the logic (1,1). The emergence of an (off-diagonal) ridge can already be seen in the right panels in Figure 5, which correspond to a relatively large value of $\beta$.

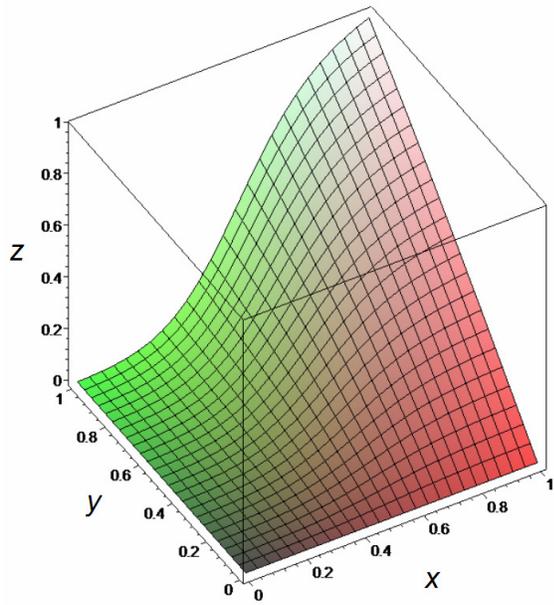 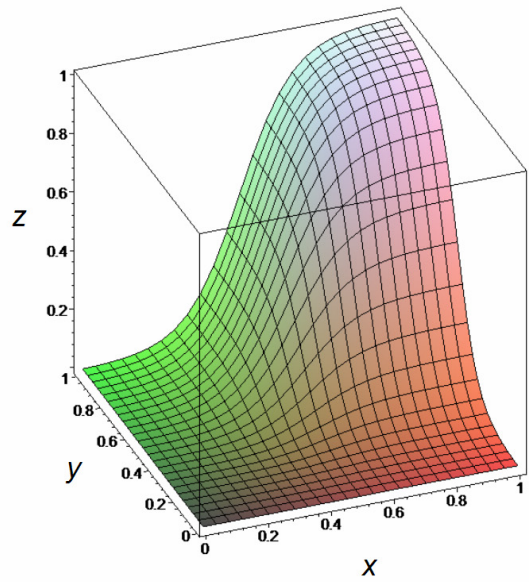

**Figure 7.** *Left:* Schematic of a one-sided sigmoid-type behavior. *Right:* A desirable two-sided sigmoid response surface for **AND** gates.

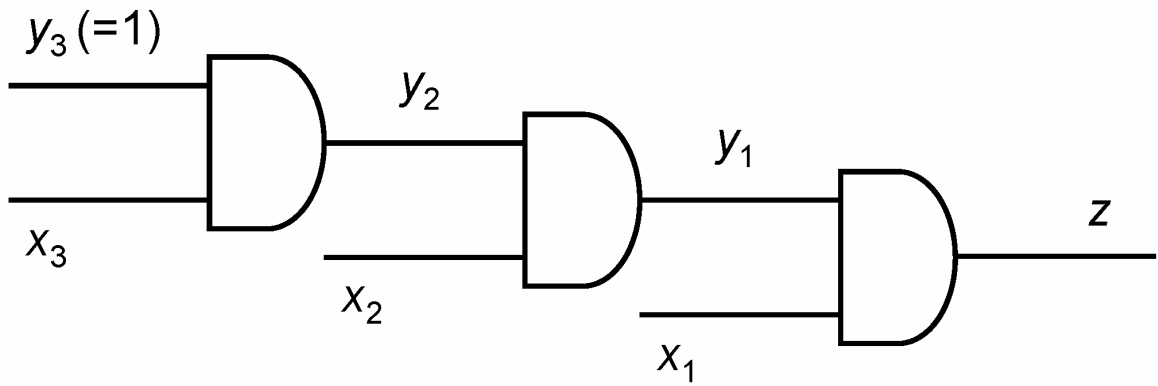

**Figure 8.** The three-**AND**-gate network, with separately varied inputs $x_{1,2,3}$ (and $y_1$ kept constant) in an experimental[19] realization.